\documentclass[9pt, twocolumn, conference, a4paper]{IEEEtran}
\usepackage[english]{babel}
\usepackage{float}
\usepackage{graphicx}
\usepackage{amsmath}
\usepackage{multirow}
\usepackage{color}
\usepackage{mathtools}
\usepackage{hyperref}

\begin{document}

\title{Deep convolutional networks for quality assessment\\ of protein folds}

\author{
\IEEEauthorblockN{
Georgy Derevyanko\IEEEauthorrefmark{1}\IEEEauthorrefmark{5},
Sergei Grudinin\IEEEauthorrefmark{2},
Yoshua Bengio\IEEEauthorrefmark{3}\IEEEauthorrefmark{4}, and
Guillaume Lamoureux\IEEEauthorrefmark{1}\IEEEauthorrefmark{6}
}
\\
\IEEEauthorblockA{
\IEEEauthorrefmark{1}
Department of Chemistry and Biochemistry
and Centre for Research in Molecular Modeling (CERMM),\\
Concordia University, Montr{\'e}al, H4B 1R6, Canada,\\
\IEEEauthorrefmark{2}
Inria, Universit{\'e} Grenoble Alpes, CNRS,
Grenoble INP, LJK, Grenoble, 38000, France,\\
\IEEEauthorrefmark{3}
Department of Computer Science and Operations Research,
Universit{\'e} de Montr{\'e}al, Montr{\'e}al, H3C 3J7, Canada\\
\IEEEauthorrefmark{4}
CIFAR Fellow\\
\IEEEauthorrefmark{5}
Email: georgy.derevyanko@gmail.com\\
\IEEEauthorrefmark{6}
Email: guillaume.lamoureux@concordia.ca\\
}
}
\maketitle

\begin{abstract}
The computational prediction of a protein
structure from its sequence generally relies on a method to assess the
quality of protein models. Most assessment methods rank candidate
models using heavily engineered structural features, defined as
complex functions of the atomic coordinates. However, very few methods
have attempted to learn these features directly from the data.
We show that deep convolutional networks can be used to predict the
ranking of model structures solely on the basis of their raw
three-dimensional atomic densities, without any feature tuning. We
develop a deep neural network that performs on par with
state-of-the-art algorithms from the literature.  The network is
trained on decoys from the CASP7 to CASP10 datasets and its
performance is tested on the CASP11 dataset.  On the CASP11 stage~2
dataset, it achieves a loss of 0.064, whereas the best performing
method achieves a loss of 0.063.  Additional testing on decoys from
the CASP12, CAMEO, and 3DRobot datasets confirms that the network
performs consistently well across a variety of protein
structures. While the network learns to assess structural decoys
globally and does not rely on any predefined features, it can be
analyzed to show that it implicitly identifies regions that deviate
from the native structure.
\end{abstract}

\section{Introduction}

The protein folding problem remains one of the outstanding challenges
in structural biology \cite{dill2012folding}.  It is usually defined
as the task of predicting the three-dimensional (3D) structure of a
protein from its amino acid sequence.
Progress in the field is monitored through the Critical Assessment of
protein Structure Prediction (CASP) competition \cite{moult1995large},
in which protein folding methods are evaluated in terms of their
accuracy at predicting structures ahead of their
publication. Most methods participating in CASP include a conformational
sampling step, which generates a number of plausible protein
conformations, and a quality assessment step, which attempts to select
the conformations closest to the unknown native structure.

In this work we explore the application of deep learning to the
problem of ``model quality assessment'' (MQA), also called
``estimation of model accuracy'' (EMA) \cite{kryshtafovych2015}. Deep
learning has recently garnered considerable interest in the
research community \cite{lecun2015deep}, particularly in computer
vision and natural language processing. Unlike more ``shallow''
machine learning approaches, deep learning improves performance by
learning a hierarchical representation of the raw data at hand. It
alleviates the need for feature engineering, which has traditionally
constituted the bulk of the work done by researchers.

Deep learning has been applied to biological data and has
yielded remarkable results for predicting the effects of genetic
variations on human RNA splicing \cite{xiong2015human}, for
identifying DNA- and RNA-binding
motifs \cite{alipanahi2015predicting}, and for predicting the effects
of non-coding DNA variants with single nucleotide
precision \cite{zhou2015predicting}. These successes have one thing in
common: they use raw data directly as input and do not attempt to
engineer features from them.

Deep-learning-inspired methods have been used for protein structure quality
assessment as well. For instance, DeepQA \cite{cao2016deepqa} uses 9
scores from other MQA methods and 7 physico-chemical features
extracted from the structure as input features to a deep restricted
Boltzmann machine \cite{hinton2006fast}. The method has been reported
to outperform ProQ2 \cite{ray2012proq2}, which was the top-performing
method in the CASP11 competition \cite{kryshtafovych2015}.  ProQ3D
\cite{uziela2017proq3d} uses the same high-level input features as
the earlier ProQ3 method \cite{uziela2016proq3} but achieves better
performance by replacing the support vector machine model by a deep
neural network. Since the original ProQ3 method had one of the top
performances in CASP12 \cite{elofsson2017qacasp12}, it can be
expected that ProQ3D performs equally well.

Although both DeepQA and ProQ3D
methods are based on deep neural networks, they use high-level
features as input. In that sense, they use deep learning models more as
traditional ``shallow'' classifiers than as end-to-end learning
models. It is likely that they do not get all the advantages offered
by the deep learning approach.
By comparison, the DL-Pro algorithm \cite{nguyen2014dlpro} uses a
sligthly more raw input, consisting of the eigenvectors of the
C$\alpha$-to-C$\alpha$ distance matrix. The model itself is an
autoencoder \cite{hinton2006reducing} trained to classify the
structures into either ``near native'' or ``not near native''.

More in line with the ``end-to-end'' spirit of deep learning, methods
using as input a 3D representation of the structure have been
developed to score protein-ligand poses \cite{wallach2015atomnet,
ragoza2017protein}, to predict ligand-binding protein
pockets \cite{jimenez2017deepsite}, and to predict the effect of a
protein mutation \cite{torng2017}. The molecules of interest are
treated as 3D objects represented on a grid and the predictions are
obtained from that information only. While a rigorous comparison of
these methods is not always possible, they appear to
improve on the state of the art: both
AtomNet \cite{wallach2015atomnet} and the 3D convolutional neural
network of \cite{ragoza2017protein} perform
consistently better than either Smina \cite{koes2013smina} or AutoDock
Vina \cite{trott2009vina}.
For small molecules, \cite{schutt2017quantum,
schutt2017moleculenet} and \cite{smith2017ani1} have recently developed deep neural networks to predict
the molecular energy of a variety of chemical compounds in various
conformations (or even various isomeric states). These models,
intended to be used as universal force fields, are trained on ab
initio quantum energies and forces, and use only the nuclear charges and the
interatomic distance matrix as input.

\section{Materials and Methods}

\subsection{Datasets}

\begin{figure*}[t]
    \makebox[\textwidth]{
    \includegraphics[width=0.78\paperwidth]{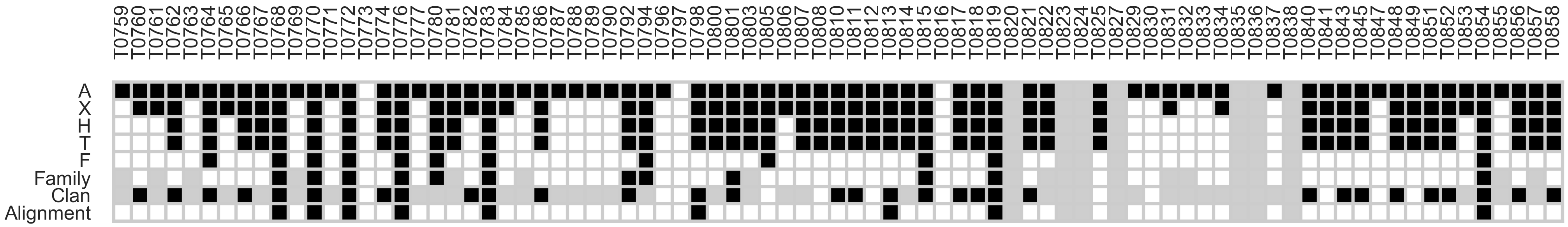}
    }
    \vspace{-15pt}
    \caption{Overlap of the training set on each target domain of the
    test set (from T0759 to T0858). The first 5 rows of tiles
    correspond to the ECOD classification of protein domains (A-, X-,
    H-, T-, and F-groups). A black tile in any of these rows indicates
    that at least one structure from the training set belongs to the
    same ECOD group as the target. A white tile indicates that no
    structure belongs to the same group. Targets for which no ECOD
    classification is available are left empty (grey).
    A black tile in the ``Family'' row indicates that at least one
    structure from the training set belongs to the same Pfam family as
    the target. (Grey indicates that no Pfam family information is
    available for the target.) The ``Clan'' row shows similar
    information for Pfam clans. A black tile in the ``Alignment'' row
    indicates that at least one sequence in the training set aligns to
    the target sequence with an E-value smaller than $10^{-4}$. (Grey
    indicates that the protein structure is absent from the PDB)}
    \label{Fig:summaryTable}
\end{figure*}

We train and assess our method using the datasets of non-native protein
conformations (``decoys'') from
the CASP competition \cite{moult2014critical}.  We use the CASP7 to
CASP10 data as training set and the CASP11 data as test set, for a
total of 564 target structures in the training set and 83 target
structures in the test set. Each target from the training set has 282
decoys on average.
The test dataset is split into two subsets \cite{kryshtafovych2015}:
``stage~1'' with 20 decoys per target selected randomly from all
server predictions and ``stage~2'' with, for each target, the 150 decoys
considered best by the Davis-QAconsensus evaluation
method \cite{kryshtafovych2015}.
The native structures are excluded from both training 
and test datasets. To make the
structural data more consistent we optimize the side chains of all decoy
structures using SCWRL4 \cite{krivov2009improved}.

Training and test datasets cover a similar range of sequence
lengths (see Figure~S1 in Supplementary Information). To confirm that the
training and test sets are significantly different, we have aligned
all test sequences against all training sequences using
blastp \cite{altschul1990basic}.  Less than 11\% of the targets in the
test set (9 out of 83) have sequence similarity with any target in the
training set (see Table~S1 in Supplementary Information).

To further assess the similarity of the two datasets, we have computed
their overlap in terms of Pfam families \cite{finn2016pfam}. 
The families were found using HMMER \cite{finn2015hmmer} with an E-value
cutoff of 1.0 \cite{finn2016pfam}.  Accounting for targets for which
no Pfam family could be determined, approximately 25\% of the test set
targets share a family with approximately 10\% of the training set
targets (see Table~S2 in Supplementary Information).

We have also compared the structures in the training and test sets
using the ECOD database \cite{cheng2014ecod}. This database provides a
5-tiered classification of all structures in the PDB
according to the following criteria:
architecture (A-group), possible homology (X-group), homology
(H-group), topology (T-group), and family (F-group).  Since the ECOD
classification is domain-based, multi-domain protein chains can belong
to multiple A-, X-, H-, T-, or F-groups.
A summary of the overlap between the training and test sets is
presented in Figure~\ref{Fig:summaryTable}. For each target domain in
the test set (T0759 to T0858), a black tile indicates that at least
one structure from the training set belongs to the same ECOD
group. (See Figure~S2 in Supplementary Information for another
representation of the overlap data.)

\begin{table}[!htb]
  \centering
  \caption {Atom types used in this work. Atoms in each group are
    identified using their standard PDB residue names and atom
    names. Asterisks (*) correspond to either 1, 2, or 3.}
\resizebox{\columnwidth}{!}{
\begin{tabular}{ c l l }
    Type & Description & Atoms \\
    \hline
    1 & Sulfur/selenium & CYS:SG, MET:SD, MSE:SE \\ \hline
    2 & Nitrogen (amide) & ASN:ND2, GLN:NE2, \\
     & & backbone N (including N-terminal) \\ \hline
    3 & Nitrogen (aromatic) & HIS:ND1/NE1, TRP:NE1 \\ \hline
    4 & Nitrogen (guanidinium) & ARG:NE/NH* \\ \hline
    5 & Nitrogen (ammonium) & LYS:NZ \\ \hline
    6 & Oxygen (carbonyl) & ASN:OD1, GLN:OE1, \\
     & & backbone O (except C-terminal) \\ \hline
    7 & Oxygen (hydroxyl) & SER:OG, THR:OG1, TYR:OH \\ \hline
    8 & Oxygen (carboxyl) & ASP:OD*, GLU:OE*, \\
     & & C-terminal O, C-terminal OXT \\ \hline
    9 & Carbon (sp2) & ARG:CZ, ASN:CG, ASP:CG, \\
     & & GLN:CD, GLU:CD, backbone C \\ \hline
    10 & Carbon (aromatic) & HIS:CG/CD2/CE1, \\
     & & PHE:CG/CD*/CE*/CZ, \\ 
     & & TRP:CG/CD*/CE*/CZ*/CH2, \\
     & & TYR:CG/CD*/CE*/CZ \\ \hline
    11 & Carbon (sp3) & ALA:CB, ARG:CB/CG/CD, \\
     & & ASN:CB, ASP:CB, CYS:CB, \\
     & & GLN:CB/CG, GLU:CB/CG, \\
     & & HIS:CB, ILE:CB/CG*/CD1, \\
     & & LEU:CB/CG/CD*, \\
     & & LYS:CB/CG/CD/CE, \\
     & & MET:CB/CG/CE, MSE:CB/CG/CE, \\
     & & PHE:CB, PRO:CB/CG/CD, \\
     & & SER:CB, THR:CB/CG2, \\
     & & TRP:CB, TYR:CB, VAL:CB/CG*, \\
     & & backbone CA \\ \hline    
\end{tabular}
}
\label{Tbl:atomTypes}
\end{table}

\subsection{Input}

\begin{figure*}[t]
    \centerline{\includegraphics[width=0.8\linewidth]{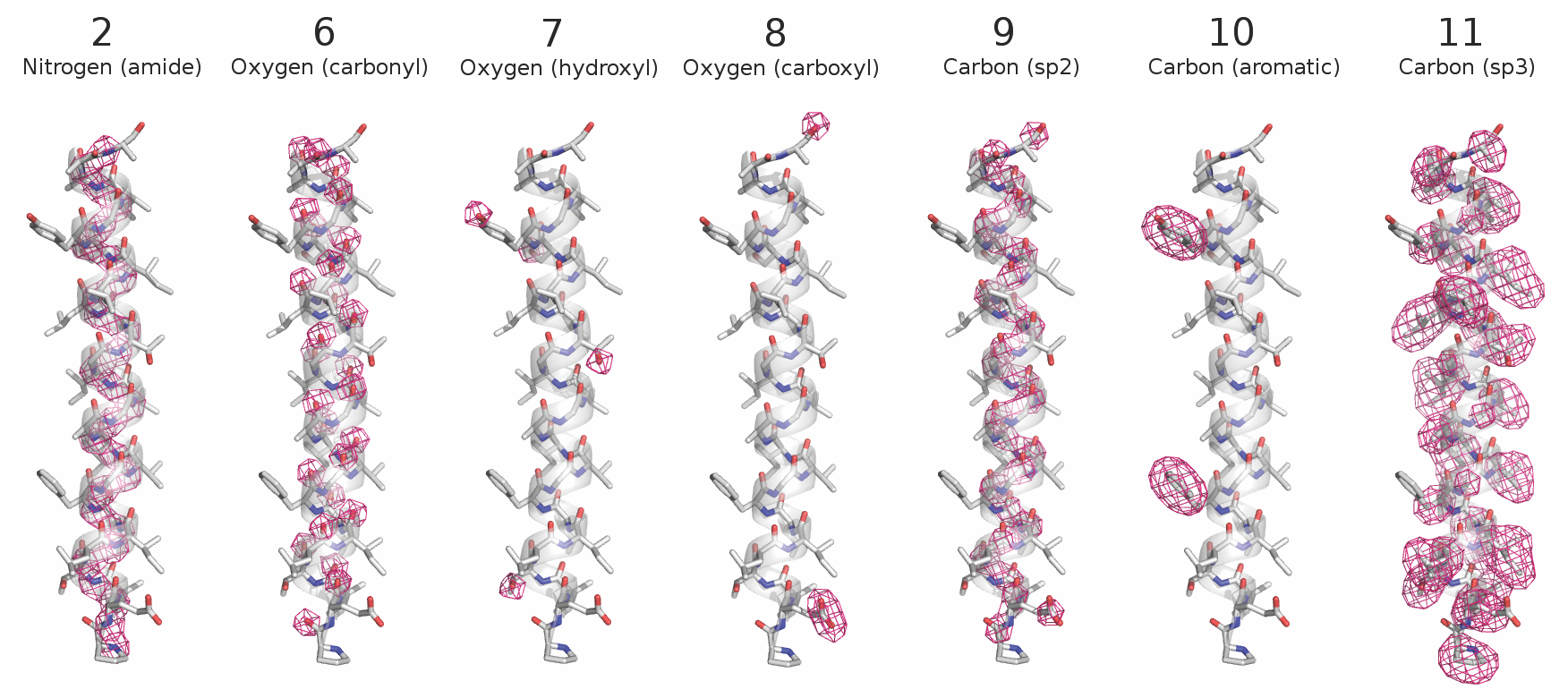}}
    \vspace{-10pt}
    \caption{Representation of a protein structure (PDB code 5eh6)
    using atomic densities. The density maps are calculated according
    to Eq.~\ref{eq:rho} and rendered using Pymol \cite{PyMOL} with an
    isosurface level of $0.5$.}
    \label{Fig:atomic_densities}
\end{figure*}

Each protein structure is represented by 11 density maps corresponding
to the atom types defined in Table~\ref{Tbl:atomTypes}. These atom
types are a simplification of the 20 types proposed by \cite{huang2006iterative, huang2008iterative}, to reduce the
memory footprint of the model.
%
%
The density of an atom is represented using the function
\begin{equation}
\rho(r) =  \begin{cases}
               e^{-\frac{r^2}{2}}&\text{if } r\leq 2.0\text{ \AA} \\
               0                 &\text{if } r>2.0\text{ \AA} \\
            \end{cases}
\label{eq:rho}
\end{equation}
The atomic density is projected to the grid corresponding to its atom
type. Each grid has a resolution of 1~\AA\ and has $120\times
120\times 120$ cells.
Figure~\ref{Fig:atomic_densities} illustrates the atomic densities for
a simple $\alpha$-helical peptide (PDB code 5eh6). This structure
contains only 7 of the 11 atom types, and only those density maps are
shown. The 4 other maps have zero density everywhere.

\subsection{Model}

\begin{figure*}[t]
    \centerline{\makebox[0.8\textwidth]{
    \includegraphics[width=\textwidth]{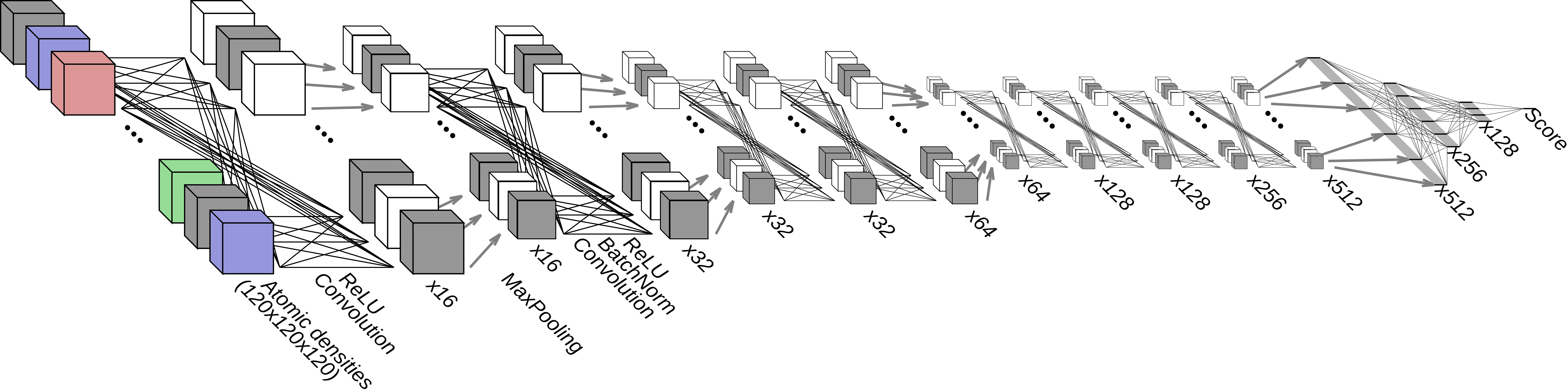}
    }}
    \vspace{-10pt}
    \caption{Schematic representation of the convolutional neural
    network architecture used in this work. Unless otherwise specified, line connections across
    boxes denote the consecutive application of a 3D convolutional
    layer (``Convolution''), a batch normalization layer
    (``BatchNorm''), and a ReLU layer. Grey arrows between boxes denote
    maximum pooling layers (``MaxPooling''). Labels ``$\times M$''
    denote the number of 3D grids and the number of filters used in the corresponding
    convolutional layer. The grey stripes denote
    one-dimensional vectors and crossed lines between them stand for
    fully-connected layers with ReLU nonlinearities. Details of the
    model can be found in Table~S3 of Supplementary Information.}
    \label{Fig:CNNModel}
\end{figure*}

In this work we score protein structures using 3D convolutional neural
networks (CNNs). CNNs were first proposed for image recognition by
\cite{lecun1989backpropagation} and first applied to biological data
by \cite{bengio1990neural}. Convolutional neural networks have gained wider
recognition after the ImageNet 2012 competition
\cite{krizhevsky2012imagenet}. The architecture of the model is shown
in Figure~\ref{Fig:CNNModel}.  It is comprised of four blocks of
alternating convolutional, batch normalization, and ReLU layers
(terminated by a maximum pooling layer), followed by three
fully-connected layers with ReLU nonlinearities. The final output of
the network is a single number, interpreted as the score of the input
structure. (See Table~S3 in Supplementary Information for more
details.)

Each 3D convolutional layer takes $N$ input density maps $f^\text{in}$
and transforms them using $M$ filters $F$ according to the following
formula:
\begin{equation}
f^\text{out}_i (\mathbf{r}) = \sum^{N}_{j=1} \int F_i (\mathbf{r} - \mathbf{r'}) \cdot f^\text{in}_j(\mathbf{r'}) ~d\mathbf{r'}, \quad\forall i \in [1,M]
\end{equation}
In practice, these convolutions are approximated by sums on a 3D grid.
The ReLU nonlinearity is computed as follows:
\begin{equation}
f^\text{out}_i (\mathbf{r}) = \begin{cases}
               f^\text{in}_i(\mathbf{r}) &\text{if } f^\text{in}_i(\mathbf{r})\geq 0\\
               0                         &\text{if } f^\text{in}_i(\mathbf{r})<0\\
            \end{cases}, \quad\forall i \in [1,M]
\end{equation}
The idea of batch normalization was introduced by 
\cite{ioffe2015batch} to reduce the shift in the distribution of
subnetwork outputs during training. This layer normalizes each input
value according to the mean and variance within the subset of examples
used to estimate the gradient (the ``batch''):
\begin{equation}
\hat{f}^\text{in}_k(\mathbf{r}) = \frac{f^\text{in}_k(\mathbf{r}) - \mu_\text{B}(\mathbf{r})}{\sqrt{\sigma^{2}_\text{B}(\mathbf{r}) + \epsilon}}, \quad\forall k \in [1,N_\text{B}]
\end{equation}
where $\mu_\text{B}(\mathbf{r})$ is the mean of all
$f^\text{in}(\mathbf{r})$ maps from the batch (calculated at each
position $\mathbf{r}$) and $\sigma^{2}_\text{B}(\mathbf{r})$ is the
variance. $N_\text{B}$ is the number of examples in the batch. The constant
$\epsilon = 10^{-5}$ is added to avoid division by zero.
The output of the layer is computed by scaling the normalized inputs:
\begin{equation}
f^\text{out}_k(\mathbf{r}) = \gamma \hat{f}^\text{in}_k(\mathbf{r}) + \beta, \quad\forall k \in [1,N_B]
\end{equation}
Parameters $\gamma$ and $\beta$ are learned along with other
parameters of the network during the training.

The maximum pooling layer (``MaxPool'') is used to build a
coarse-grained representation of the input. The output of this layer
is the maximum over the cubes of size $d \times d \times d$ that cover
the input domain with a stride $l$ in each direction.  This operation
makes the output size approximately $l$ times smaller than the
input in each direction.  All four ``MaxPool'' layers of the model
(Figure~\ref{Fig:CNNModel}) use $d=3$ and $l=2$.

During the coarse-graining procedure, the size of the individual data
grids eventually shrinks to a single cell. The flattening layer
reshapes the array of $1\times 1\times 1$ density maps into a single
vector. Afterwards, we compute several transformations using
fully-connected layers. Each of these layers transform a vector
$\mathbf{x}_\text{in}$ as follows:
\begin{equation}
\mathbf{x}_\text{out} = W \cdot \mathbf{x}_\text{in} + \mathbf{b}
\end{equation}
where $W$ is a rectangular matrix and $\mathbf{b}$ is a vector,
learned during the training. Each output vector is then transformed by
a ReLU layer.

\subsection{Training loss function}

The problem of decoy quality assessment is essentially a ranking
problem: we have to arrange decoys according to their similarity to
the corresponding native structure as quantified, for instance, by the
GDT\_TS score \cite{zemla2001casp4}. Such a ranking approach has
recently been used by the MQAPRank method \cite{jing2016sorting},
which, however, relies on a support vector machine model and uses
high-level features as input.

We define the training loss function in terms of
the margin ranking loss \cite{joachims2002optimizing, gong2013deep}
for each pair of decoys.  Let $\text{GDT\_TS}_i$ denote the global
distance test total score of decoy $i$ and let $y_{ij}$ be the
ordering coefficient of two decoys $i$ and $j$:
\begin{equation}
y_{ij} = \begin{cases}
               1& \text{if }\text{GDT\_TS}_i \leq \text{GDT\_TS}_j \\
               -1& \text{if }\text{GDT\_TS}_i > \text{GDT\_TS}_j \\
            \end{cases}
\end{equation}
The original GDT\_TS score covers the range $[0,100]$ but in this work
we use a GDT\_TS score normalized to the range $[0,1]$, so that the
loss stays within reasonable bounds.  Let $s_i$ denote the output of
the network for decoy $i$. We use the following expression for the
pairwise ranking loss:
\begin{equation}
L_{ij} = w_{ij} \max \left[ 0, 1 - y_{ij} \cdot (s_i - s_j) \right]
\end{equation}
The coefficient $w_{ij}$ represents the weight of each example and is
defined so that decoys with similar scores (within $0.1$) are removed
from the training:
\begin{equation}
w_{ij} = \begin{cases}
               1& \text{if } \left| \text{GDT\_TS}_i - \text{GDT\_TS}_j \right| > 0.1 \\
               0& \text{otherwise} \\ 
            \end{cases}
\end{equation}

During the training procedure we load $N_\text{B}$ decoy structures of
a given target into memory (a ``batch'') and compute the output of the
network and the average ranking loss:
\begin{equation}
L = \frac{1}{N_\text{B}^2} \sum_{i=1}^{N_\text{B}}\sum_{j=1}^{N_\text{B}} L_{ij}
\label{Eq:loss}
\end{equation}
In principle, the ranking loss \ref{Eq:loss} would be minimal for any
output $s$ that decreases monotonically with GDT\_TS. While an output
$s$ strictly equal to the negative of the GDT\_TS score would produce
a loss of zero, our preliminary experiments have shown better
performance when the model is trained so that $s$ orders like the
negative of GDT\_TS without necessarily being equal to it.

\subsection{Evaluation criteria}

We evaluate the model using various correlation coefficients of the
scores and using a evaluation loss function distinct from the training loss
function. The evaluation loss is defined, for any given protein, as the
absolute difference between the GDT\_TS of the best decoy and the
GDT\_TS of the decoy with the lowest predicted score $s$:
\begin{equation}
\mathrm{Loss} = \left| \mathrm{max}_i(\text{GDT\_TS}_i) - \text{GDT\_TS}_{\mathrm{argmin}_i(s_i)} \right|
\label{Eq:eloss}
\end{equation}
The correlation coefficients between the $s$ score produced by the
model and the GDT\_TS score are computed for all decoys of a given
target in the test set and are then averaged over all targets. Since
the value of GDT\_TS increases with the quality of a model but the
value of $s$ decreases, an ideal MQA algorithm would show the correlation
coefficient of $-1$ and zero loss.
These two evaluation criteria measure different qualities of the
model. One the one hand, a perfect correlation coefficient of $-1$ would
be achieved if the algorithm ranks all decoys in the exact
order of their GDT\_TS score (from best to worst). On the other hand,
a zero loss would be achieved if the algorithm systematically assigns
the lowest $s$ value to the decoy with the highest GDT\_TS score,
irrespective of the $s$ value it assigns to the other decoys.

\subsection{Optimization and dataset sampling}

The parameter optimization of the model was performed using the Adam
algorithm \cite{kingma2014adam}. The gradient of the 
average training loss function (Eq.~\ref{Eq:loss})
with respect to the model parameters is computed on the pairs of
models in the batch. The batch size was set to $N_\text{B} = 9$
models.

The training dataset is sampled by first choosing a random target from
the dataset, then sampling decoys of this target. One epoch
corresponds to one pass through all targets in the dataset. The decoys
are sampled in a homogeneous way, by dividing all decoys of a given
target into $N_\text{B}$ bins according to the value of their GDT\_TS score
and by picking one decoy from each bin at random.
Precisely, decoy $i$ belongs to bin number 
\begin{equation}
1 + \left\lfloor{ N_B \times \frac{\text{GDT\_TS}_i - \min(\text{GDT\_TS}) }{\max(\text{GDT\_TS}) - \min(\text{GDT\_TS})} }\right\rfloor
\end{equation}
where $\max(\text{GDT\_TS})$ and $\min(\text{GDT\_TS})$ are computed
on all decoys of the chosen target.  If a bin
is empty, the decoy is picked from another non-empty bin chosen at
random.  The order of targets and the order of decoys in the bins are
shuffled at the end of each epoch.

Decoy structures are randomly rotated and translated each time they
are used as input. The rotations are sampled uniformly
\cite{shoemake1992uniform} and the translation are chosen in such a
way that the translated protein fits inside the $120$~\AA${}\times
120$~\AA${}\times 120$~\AA\ input grid (see text in Supplementary
Information for details).

\begin{figure}[!tpb]
    \centerline{\includegraphics[width=0.7\linewidth]{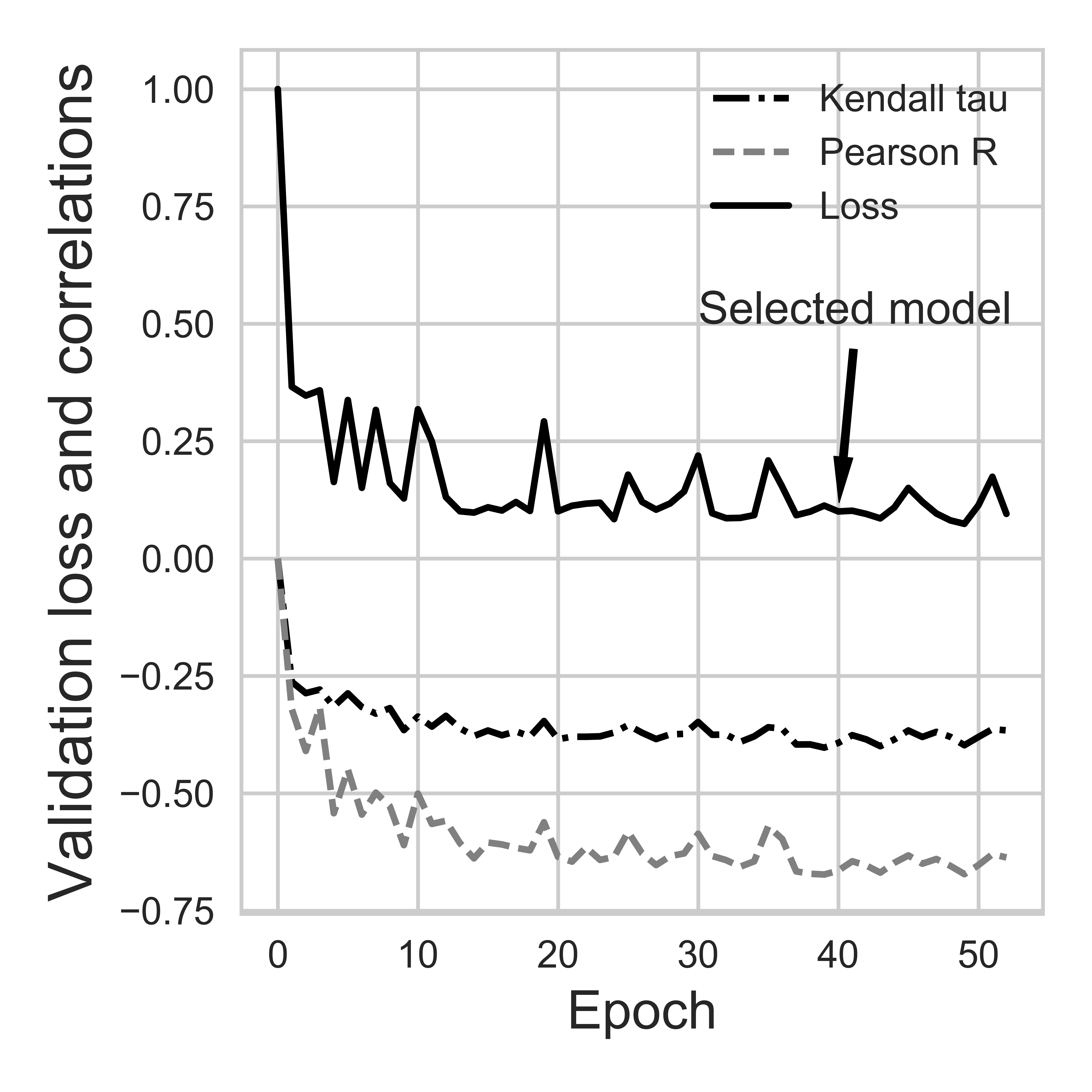}}
    \vspace{-15pt}
    \caption{Evaluation loss (Eq.~\ref{Eq:eloss}), Kendall $\tau$, and Pearson $R$ coefficients
      evaluated on the validation subset during the training
      procedure.  One epoch corresponds to a cycle over all targets in
      the training subset. Models are saved every 10 epochs and the
      arrow shows the minimum validation loss for which a model was
      saved (at epoch 40).}
    \label{Fig:TrainingLoss}
\end{figure}

We select the final model based on its performance on a
\emph{validation subset} consisting of 35 targets (and their decoys)
picked at random from the training set and excluded from the training
procedure. The remaining 529 targets are called the \emph{training
subset}.  Figure~\ref{Fig:TrainingLoss} shows the Kendall $\tau$ and
Pearson $R$ coefficients and the evaluation loss on the validation subset over 52
epochs of training.  Models are saved every 10 epochs and we pick the
one that has the smallest evaluation loss (at epoch 40).
Table~\ref{Tbl:TrainingResults} summarizes the performance metrics on
the training and validation sets for the model at epoch 40.
(See Figure~S3 for results broken down by target.)

\section{Results}

\begin{figure}[t]
    \centerline{\includegraphics[width=0.7\linewidth]{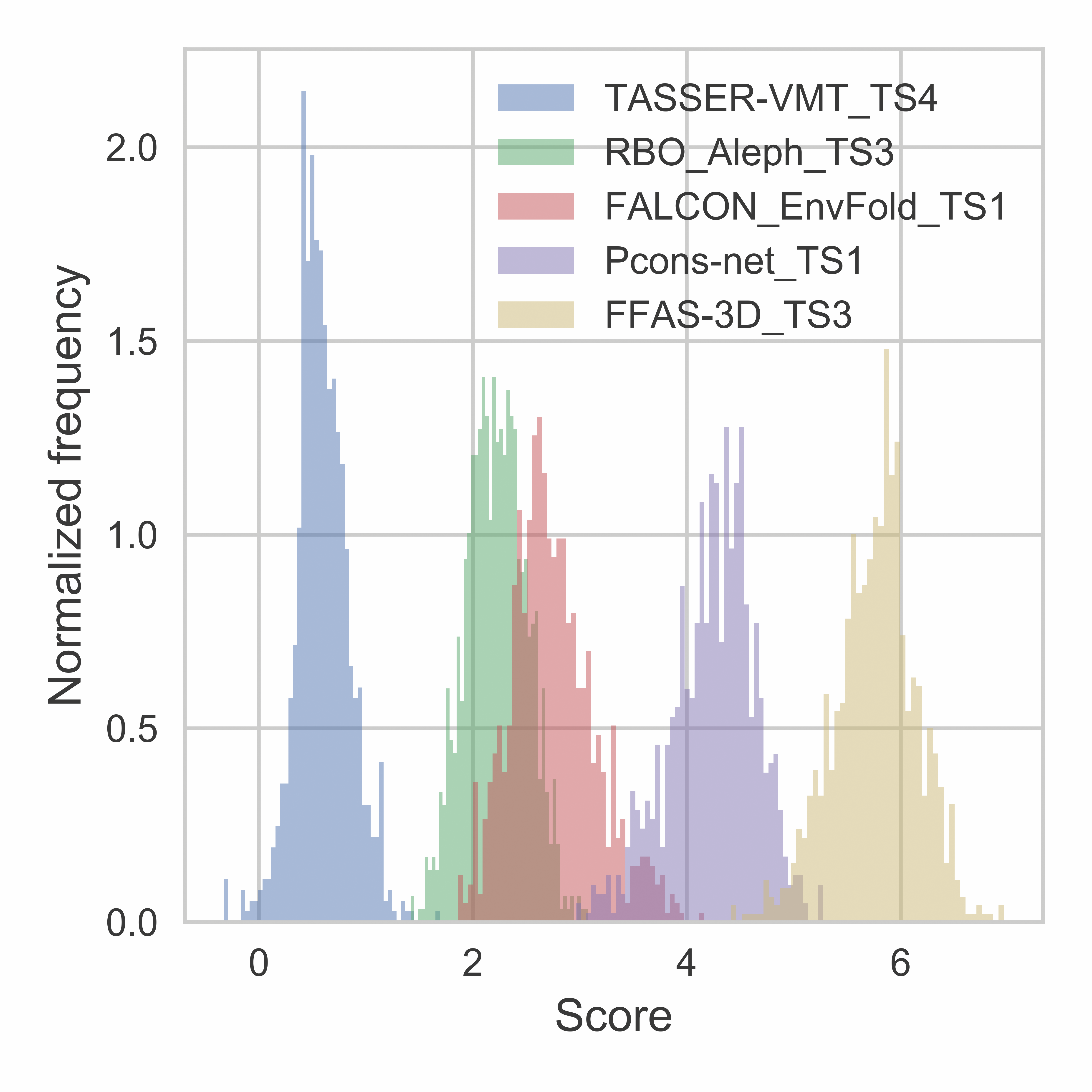}}
    \vspace{-15pt}
    \caption{Distributions of the $s$ scores of five decoys for target
    T0832 under random translations and rotations. A lower score
    represents a higher quality.}
    \label{Fig:DecoysScoreDistribution}
\end{figure}

\begin{table}[!t]
  \centering
  \caption {Performance of the 3DCNN model from epoch 40 on the training and
    validation subsets.}
\resizebox{\columnwidth}{!}{
\begin{tabular}{ c c c c c }
    Data & Loss (Eq.~\ref{Eq:eloss}) & Pearson $R$ & Spearman $\rho$ & Kendall $\tau$ \\
    \hline
    Training subset     &0.146 &0.71 &0.61 &0.45 \\
    Validation subset   &0.135 &0.71 &0.59 &0.44 \\ \hline
\end{tabular}
}
\label{Tbl:TrainingResults}
\end{table}
Ideally, the score assigned to a decoy should not depend on its
position and orientation in space. To allow the model to learn this
invariance, the rotational and translational degrees of freedom of all
decoy structures are randomly sampled during the training.
Figure~\ref{Fig:DecoysScoreDistribution} shows the distributions of
scores for several decoy structures of the same target (T0832),
calculated using the trained model for 900 rotations and translations
sampled uniformly.  While the score of a given structure is not
strictly invariant under rotation and translation, it has a relatively
narrow, unimodal distribution.  (See Figure~S4 in Supplementary
Information for a distribution of score under rotations and
translations separately.)  More importantly, the difference between
the average scores of two decoys is usually larger than their
standard deviations.
To reduce the influence of the choice of rotation and translation on
the final ranking, we estimate the score of each decoy from the
average of 90 scores calculated for random rotations and translations.

\subsection{Performance on the CASP11 benchmark}

Table~\ref{Tbl:TestResults} shows a comparison of our model (3DCNN)
with a number of state-of-the-art MQA methods: ProQ2D,
ProQ3D~\cite{uziela2017proq3d},
VoroMQA~\cite{olechnovivc2017voromqa}, and
RWplus~\cite{zhang2010novel}.
(See Figures~S5 and S6 for ranking results broken down by target.)
ProQ2D uses a number of carefully crafted features such as atomic
contacts, residue-residue contacts, surface accessibilities (as found
in the structure and as predicted from the sequence), and secondary
structure (observed and predicted). ProQ3D employs the same features
as ProQ2D, as well as some Rosetta energy
terms~\cite{leaverfay2011rosetta}.
RWplus, similar to DOPE~\cite{shen2006statistical} and
DFIRE~\cite{zhou2002distance}, uses a scoring approach based on
statistical pairwise potentials.
\begin{table}[t]
  \centering
  \caption{Performance comparison of our method (3DCNN) with other
    state-of-the-art MQA methods on the CASP11 dataset stages~1 and 2
    (see text). The table reports the absolute, per-target average
    values of the correlation coefficients.}
\resizebox{\columnwidth}{!}{
\begin{tabular}{ c c c c c }
    MQA method & Loss (Eq.~\ref{Eq:eloss}) & Pearson $R$ & Spearmann $\rho$ & Kendall $\tau$ \\ \hline
    \multicolumn{5}{ l }{Stage 1} \\ \hline
    ProQ3D   &0.046 &0.755 &0.673 &0.529 \\
    ProQ2D   &0.064 &0.729 &0.604 &0.468 \\
    \textbf{3DCNN} &0.064 &0.535 &0.425 &0.325 \\    
    VoroMQA  &0.087 &0.637 &0.521 &0.394 \\
    RWplus   &0.122 &0.512 &0.402 &0.303 \\ \hline    
    \multicolumn{5}{ l }{Stage 2} \\ \hline
    VoroMQA  &0.063 &0.457 &0.449 &0.321 \\ 
    \textbf{3DCNN} &0.064 &0.421 &0.409 &0.288 \\
    ProQ3D   &0.066 &0.452 &0.433 &0.307 \\
    ProQ2D   &0.072 &0.437 &0.422 &0.299 \\
    RWplus   &0.089 &0.206 &0.248 &0.176 \\ \hline
\end{tabular}
}
\label{Tbl:TestResults}
\end{table}

VoroMQA uses knowledge-based potentials that depend on the contact
surface between pairs of heavy atoms in the protein (or the
solvent). Its approach is distinct from both the machine-learning
techniques exemplified by the ``ProQ'' methods and the statistical
potential techniques exemplified by the RWplus method.
The methods chosen have available codes and could be re-evaluated on
our CASP11 benchmark. Targets T0797, T0798, T0825 were removed from
the benchmark because they were released for multimeric prediction.
All methods were re-evaluated using the default settings.

Methods ProQ2D and ProQ3D are trained on the CASP9 and CASP10 models
\cite{uziela2017proq3d}, using features trained on a diverse set of
protein structures \cite{ray2012proq2, uziela2016proq3}.  The VoroMQA
method is trained on high-resolution, nonredundant structures from the
PDB \cite{olechnovivc2017voromqa} (2.5~\AA\ resolution cutoff and 50\%
sequence identity cutoff, for a total of 12,825 PDB entries).  The
RWplus scoring function is trained on the CASP7 and CASP8 models
\cite{zhang2010novel}, using a statistical potential trained on
high-resolution structures from the PDB (1.6~\AA\ resolution cutoff and
20\% sequence identity cutoff, for a total of 1,383 PDB entries).

\begin{figure}[t]
    \centerline{\includegraphics[width=0.9\linewidth]{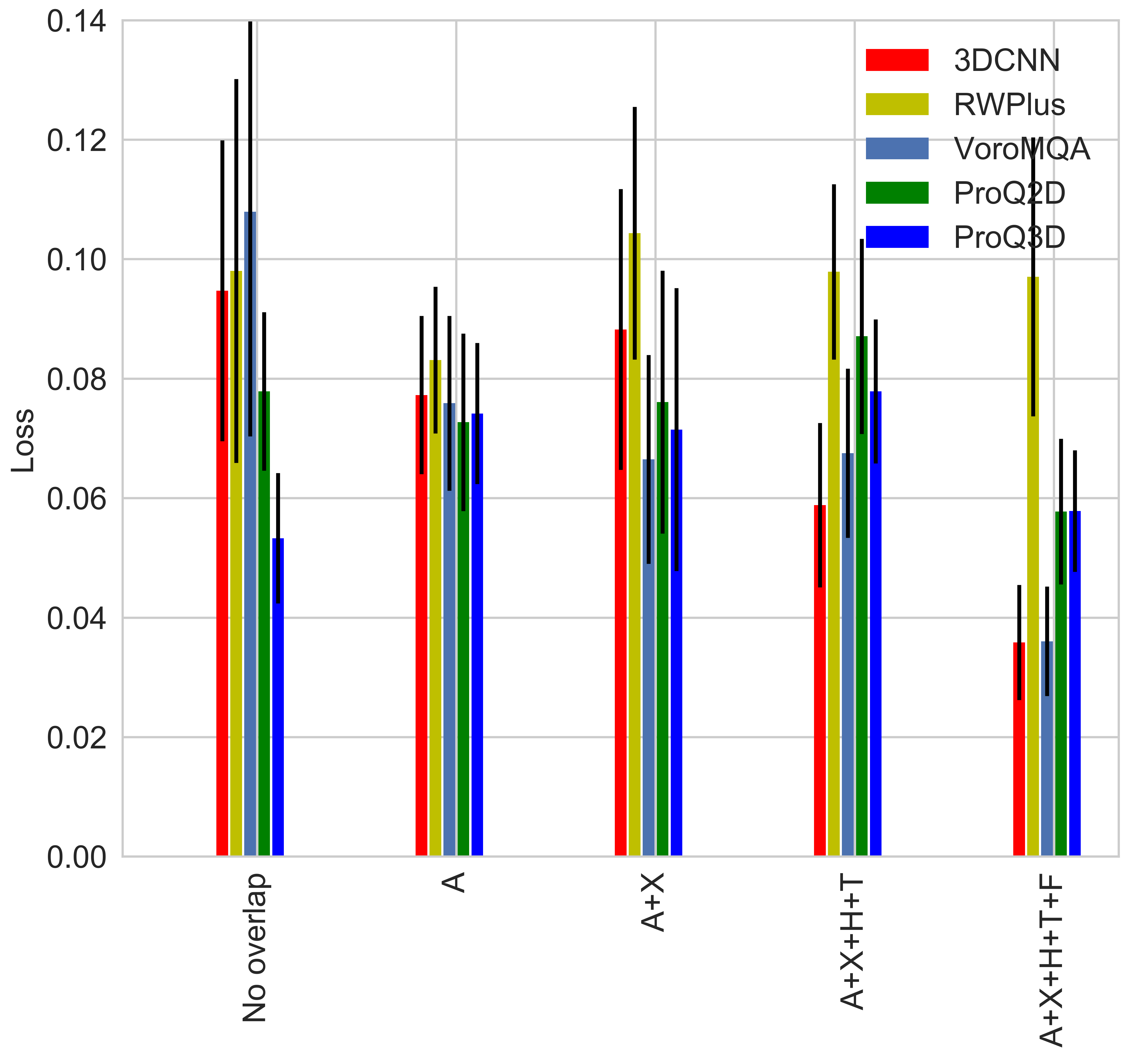}}
    \vspace{-10pt}
    \caption{Per-target average loss of the MQA algorithms of Table~3
      on the CASP11 test set stage~2, divided into 5 subsets of
      increasing structural similarity with the training set. The
      subsets are chosen according to the presence in the training set
      of structures belonging to the same ECOD categories (see
      Figure~1).
    ``No overlap'' represent the structures which have no ECOD group
      in common with the training set (T0797 and T0773);
    ``A'', the structures in the same A-group of at least one training
      structure but not in the same X-group (T0759, T0763, T0769,
      etc.);
    ``A+X'', the structures in the same X-group of at least one
      training structure but not in the same H-group (T0760, T0761,
      T0765, etc.);
    ``A+X+H+T'', the structures in the same T-group of at least one
      training structure but not in the same F-group (T0762, T0766,
      T0767, etc.); and
    ``A+X+H+T+F'', the structures in the same F-group of at least one
      training structure (T0764, T0768, T0770, etc.).
    Error bars show per-target standard error of the mean.}
    \label{Fig:Loss_vs_ECOD}
\end{figure}

Despite relying solely on atomic coordinates, the 3DCNN model (Figure~\ref{Fig:CNNModel}) achieves
a performance comparable to those of the heavily engineered ProQ2D and
ProQ3D models, with evaluation losses either slightly above or
slightly below, depending on the test set. 
Figure~\ref{Fig:Loss_vs_ECOD} shows that the performance of the 3DCNN
model increases as structural similarity with the training set
increases. This suggests that the model relies on memorization and
possibly overfits some of the structures more similar to the training
set. Surprisingly, the same trend can be seen for VoroMQA, which is a
scoring method based on local atomic packing
\cite{olechnovivc2017voromqa}. By contrast, the performances of RWPlus
(trained on CASP7 and CASP8 data) and of ProQ2D and ProQ3D (trained on
CASP9 and CASP10 data) appear less sensitive to the level of
structural similarity.

\begin{figure*}[!t]
    \centerline{\includegraphics[width=0.8\linewidth]{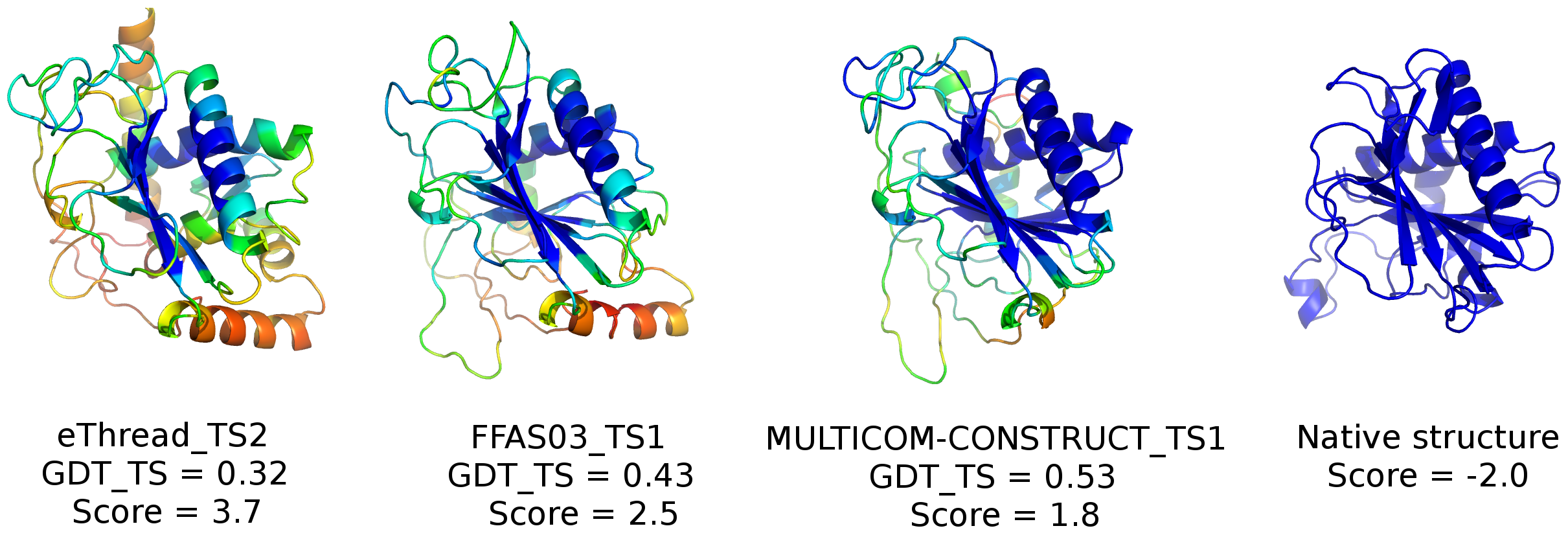}}
    \vspace{-10pt}
    \caption{Output of the Grad-CAM analysis for layer 10 of the
      network projected on four decoys of target T0786 (PDB code
      4QVU).  The values are represented using the ``rainbow'' color
      scheme.  The decoys are arranged in increasing quality from left
      to right, with the native structure on the right.  Each decoy is
      aligned on the native structure and viewed in the same
      orientation.}
    \label{Fig:GradCAMT0786_more}
\end{figure*}

\subsection{Analysis}

In this section we show that the 3DCNN network has learned a relevant
description of the protein structure and not merely artifacts of the
dataset that correlate with the desired outcome.

First, we identify the regions of a decoy structure that are
responsible for an increase of its score (a \emph{decrease} in its
quality). If the network has learned interpretable features of the
input, we expect these parts of the decoy to deviate from the native
structure.
We use the Grad-CAM analysis technique proposed by \cite{selvaraju2016grad}. 
The key idea of this technique is to
compute the gradient of the final score with respect to the output of
a certain layer of the network, then compute the sum of this layer
output weighted by the gradient.
The weighted sum highlights the regions of the layer that are both
strongly activated and highly influential on the final score. To
generate an interpretable map, the weighted sum is then scaled up to
the size of the input of the network, using tri-linear interpolation.
This up-sampled map indicates which parts of the input contribute
the most to the gradient of the score.  In our case we choose to
analyze layer 10, for which the output grid size is $25\times 25\times
25$.  We tested the method on neighboring layers and
layer 10 represents the best tradeoff between interpretability and
coarseness.
In line with our scoring procedure, we average the results
from the Grad-CAM analysis over 90 rotations and translations of the
decoy. We obtain the Grad-CAM output for each transformation and
project it onto the atoms of the decoy.  

Figure~\ref{Fig:GradCAMT0786_more} shows a projection of the Grad-CAM
results onto the atoms of four decoys of target T0786, represented as
a color-coded value on the cartoon rendering of the structures. The
orange/yellow regions are mainly found at the surface of the
lower-quality decoys while the blue/green regions are found at the
core. This indicates that the quality of the decoy would go down for
any increase in atomic density at the surface but would be unaffected
by an increase in density at the core (see Figure~S7 in Supplementary Information). It also suggests that the
neural network recognizes and enforces packing.
Moreover, we see that the higher secondary structure content of decoy
eThread\_TS2 actually decreases the score.  This suggests that the
3DCNN network does not estimate the quality based only on the presence
of local structural elements but that it detects large-scale features
of the fold as well.
Interestingly, we find that the Grad-CAM outputs are mostly zero for
decoys close to the native structure, despite the fact that no
gradient information was included in the training procedure (see
Table~S4 in Supplementary Information).

To verify that the network does not rely on artifacts in the data to
rank decoys, we have assessed its performance on three additional
independent datasets (see Table~S5 in Supplementary information).
The ``CASP12'' dataset contains all decoys from the CASP12 competition
\cite{elofsson2017qacasp12} available as of December 2, 2017. On this
dataset, the pre-trained 3DCNN model yields displays an evaluation
loss smaller than all other models tested: $0.146$, compared to
$0.151$ for ProQ2D, $0.161$ for VoroMQA, $0.161$ for ProQ3D, and
$0.192$ for RWplus (see Table~S5).
The ``CAMEO'' dataset contains all structural models published on the
CAMEO-QE webpage \cite{haas2013cameo} in the 6-month period prior to
December 10, 2017. On this dataset, the 3DCNN model performs
significantly better than both VoroMQA and RWplus, the two other
models tested (see Table~S5).
The ``3DRobot'' dataset, generated by the 3DRobot algorithm
\cite{deng20163drobot}, consists of 300 decoys for each of 200
single-domain proteins selected from the PDB. The proteins have less
than 20\% sequence identity with one another and are between 80 to 250
residue long. The algorithm yields decoys that are uniformly
distributed within an RMSD range of 0 to 12~\AA\ away from the native
structure. Out of the 200 proteins, 48 are all-$\alpha$ proteins, 40
are all-$\beta$, and 112 are $\alpha/\beta$.
The evaluation loss for these 60000 decoy structures is larger for the
3DCNN model than for VoroMQA and RWplus. However, all three models
yield high correlation coefficients. The pre-trained 3DCNN network
gives an absolute per-target average Pearson $R$ coefficient of
$0.856$, compared to $0.891$ for VoroMQA and $0.844$ for RWplus. The
Spearman $\rho$ and Kendall $\tau$ coefficient are $0.839$ and
$0.652$, respectively (see Table~S5; see Figure~S8 for representative
examples of score versus GDT\_TS plots).
This confirms that the 3DCNN model can successfully rank unrelated
datasets.

\section{Discussion}

This work shows that it is possible to construct an algorithm that
learns to assess the quality of protein models from a raw
representation. Here, we have used 3D atomic densities broken down by atom
types. However it is clear that any other
physical quantity defined on a grid can be employed, such as the
electrostatic potential calculated using the Poisson-Boltzmann equation
\cite{honig95} or the solvent density calculated using 3D-RISM
\cite{stumpe2011}. So far, no other MQA method has managed
to include these crucial properties.

The loss function we used for training does not aim to predict GDT\_TS
of a decoy, but rather to sort decoys according to their quality. We
chose this strategy so that the score can be interpreted as an energy
function that has a local minimum for the native structure.
In future
work we plan to add terms to the loss function that penalize the first
and second order derivatives of the loss at the native structure, to
ensure that the score indeed reaches a local minimum there.

This work also identifies important avenues for improvement. First,
the model captures the invariance of the score under translations and
rotations only in an approximate way. This invariance problem can
however be solved using the approach of \cite{worrall2016harmonic},
in which the coefficient space of
the convolutional filters is restricted to circular harmonics, which
encodes equivariance under rotations at each layer of the network and
leads to invariance of the final output.
Second, the output of the model remains difficult to interpret. While
interpretation of deep neural networks remains an important research
problem, the field is undergoing rapid progress. For instance,
recently published work \cite{bau2017network} has shown that
intepretability can be quantified using extensively labeled image
datasets that contain the bounding boxes and labels for fine-grained
features such as body parts or car parts. In the case of protein
models, many such labels (and bounding boxes) are readily available:
amino acids, secondary structure elements, hydrogen bond networks,
disulfide bonds, etc. Unlike in conventional machine learning models,
these features would not be used for prediction but for interpretation
of the prediction.

\section*{Supplementary information}
Supplementary data are available at
\href{https://github.com/lamoureux-lab/3DCNN_MQA/raw/Release/doc/SI.pdf}{https://github.com/lamoureux-lab/3DCNN\_MQA/raw/Release/doc/SI.pdf}.
The code and the datasets are available at
\href{https://github.com/lamoureux-lab/3DCNN_MQA}{https://github.com/lamoureux-lab/3DCNN\_MQA}.

\section*{Acknowledgments}
This work was supported by the Natural Sciences and Engineering
Research Council of Canada (NSERC) (RGPIN 355789 to G.L. and RGPIN
1016552 to Y.B.) and the Canada Research Chair and Canadian
Institute for Advanced Research (CIFAR) programs (to Y.B.).
Computational resources were provided by Calcul Qu{\'e}bec and Compute
Canada.

\bibliography{citations}
\bibliographystyle{unsrt}

\end{document}